\documentclass[aps,prb,superscriptaddress,twocolumn,showpacs]{revtex4}
\usepackage{amsmath}
\usepackage{bm}
\usepackage{graphicx,epsfig}

\renewcommand{\v}[1]{{\bf #1}}
\newcommand{\be}{\begin{equation}}
\newcommand{\ee}{\end{equation}}
\newcommand{\bd}{\begin{displaymath}}
\newcommand{\ed}{\end{displaymath}}
\newcommand{\ba}{\begin{eqnarray}}
\newcommand{\ea}{\end{eqnarray}}
\newcommand{\nn}{\nonumber \\}
\newcommand{\bpm}{\begin{pmatrix}}
\newcommand{\epm}{\end{pmatrix}}

\begin{document}

\title{Skyrmion Generation by Current}

\author{Youngbin Tchoe}
\affiliation{Department of Physics and Astronomy, Seoul National
University, Seoul 151-742, Korea}
\author{Jung Hoon Han}
\email[Electronic address:$~~$]{hanjh@skku.edu}
\affiliation{Department of Physics and BK21 Physics Research
Division, Sungkyunkwan University, Suwon 440-746, Korea}
\affiliation{Asia Pacific Center for Theoretical Physics, POSTECH,
Pohang, Gyeongbuk 790-784, Korea}

%\address{ }
\date{\today}

\begin{abstract} Skyrmions, once a hypothesized field-theoretical
object believed to describe the nature of elementary particles,
became common sightings in recent years among several
non-centrosymmetric metallic ferromagnets. For more practical
applications of Skyrmionic matter as carriers of information, thus
realizing the prospect of ``Skyrmionics", it is necessary to have
the means to create and manipulate Skyrmions individually. We show
through extensive simulation of the Landau-Lifshitz-Gilbert equation
that a circulating current imparted to the metallic chiral
ferromagnetic system can create isolated Skyrmionic spin texture
without the aid of external magnetic field.
\end{abstract}
\pacs{73.43.Cd, 72.25.-b, 72.80.-r}
\maketitle

\section{Introduction}
\label{sec:intro}

Skyrmions are topological configurations of a vector order parameter
in field-theoretical systems, whose existence had been anticipated
mathematically by the particle physicist Tony Skyrme for half a
century\cite{skyrme}. The two-dimensional version of it, sometimes
known as the baby Skyrmion, was recognized to be the topological
solution of the non-linear sigma model, which serves as a model for
ferromagnets in two dimensions\cite{rajaraman}. The discovery of
two-dimensional Skyrmions took place in a variety of condensed
matter systems such as quantum Hall
ferromagnets\cite{QH-skyrmion-theory,QH-skyrmion-exp}, metallic
chiral ferromagnets\cite{pfleiderer1,pfleiderer2,tokura,FeGe}, as
well as the ferromagnetic monolayer\cite{blugel} and doped
antiferromanget\cite{antiferro}, and is predicted to be possible in
two-dimensional liquid crystal systems\cite{liquid-crystal}. A
series of metallic magnets of B20 structure including
MnSi\cite{pfleiderer1}, Fe$_{1-x}$Co$_x$Si\cite{pfleiderer2,tokura},
and FeGe\cite{FeGe} is now known to host the Skyrmionic magnetic
texture. In these magnetic materials the formation of Skyrmions is a
consequence of the competitive interplay between the
Dzyaloshinskii-Moriya (DM) interaction which tends to induce
magnetic spirals, and an external magnetic field which tends to
bring about a ferromagnetic background\cite{bogdanov,han}. Initial
theoretical consideration assumed a layered structure of Skyrmions
embedded in a three-dimensional crystal and found a small region for
its stability just below the magnetic ordering transition
temperature\cite{bogdanov,pfleiderer1}. Later, Monte Carlo
simulation for two-dimensional magnetic model showed a much wider
region for its existence extending down nearly to zero
temperature\cite{han2,tokura}, as was indeed verified in experiment
on a thin-film Fe$_{0.5}$Co$_{0.5}$Si\cite{tokura}, due to the lack
of competing magnetic structures in the thin-film geometry. Upon
increasing the thickness of the film, the Skyrmion phase gradually
moves to a higher temperature regime\cite{FeGe}.

Aside from their fundamental scientific allure, it is of importance
now to ask if these novel topological objects can be manipulated to
the extent that they can be used as carriers of charge, spin, and of
information in general. Applying external magnetic field to a chiral
ferromagnet generates a lattice of Skyrmions, or a Skyrmion
crystal\cite{pfleiderer1,pfleiderer2,tokura,FeGe}. Spin wave
fluctuations can be induced in the Skyrmion crystal by applying an
additional ac magnetic field as was recently studied
theoretically\cite{tchernyshyov,mochizuki}. Other than by magnetic
field, the Skyrmions can be manipulated through the Hund's rule
coupling of the spin of electrons in the metallic host to the
localized moments forming the Skyrmionic texture. Recent
demonstration of the rotation of the Skyrmion crystal axis by the
applied current\cite{torque,rosch} is an example of this kind. In
the absence of pinning, Skyrmions are expected to move with the
velocity equal to the drift velocity of the electrons responsible
for the current\cite{zang}. Gilbert damping modifies the Skyrmion
trajectory in such a way to induce Hall-like motion of Skyrmion
orthogonal to the current direction\cite{zang}.

Although various aspects of current-induced motion of Skyrmions have
been studied theoretically in a number of papers
already\cite{rosch,zang}, the possibility of \textit{generating
Skyrmions by the current} has not been addressed yet. From a
technological point of view, it is desirable to have the extra means
to generate and destroy Skyrmions one at a time, instead of having
to generate them only in the form of a crystal by the external
field. Such possibility, once realized, will help usher the era of
``Skyrmionics" - an effort to tailor Skyrmions as carriers of
information bits. In this article, using extensive simulation of
Landau-Lifshitz-Gilbert (LLG) equation, we provide evidence that
circulating spin current can couple to local magnetic moments to
produce Skyrmion spin texture in the latter. Basic theory and LLG
equation are introduced in Sec. \ref{sec:formulation}. Numerical
findings are given in Sec. \ref{sec:process}. Possibility of
experimental realization will be discussed in Sec. \ref{sec:exp}.

\section{Formulation}
\label{sec:formulation}

We imagine a thin slab of chiral ferromagnet, such that all degrees
of freedom behave identically along the thickness direction. The
dynamics of the localized moments $\v n$ are governed by the
Hamiltonian\cite{tokura,zang}

\ba H_{\v n} &=& {J\over 2a} \int d^3 \v r ~ \sum_{\mu=x,y}
(\partial_\mu \v n) \cdot (\partial_\mu \v n) \nn
 && ~~~ +\! {D\over a^2} \int d^3 \v r ~
\v n \cdot \bm \nabla \times \v n  - {1\over a^3} \int d^3 \v r ~ \v
B \cdot \v n . \label{eq:H}\ea
Formulated in the continuum, $a$ expresses the linear dimension of
the suitably chosen unit cell, while $J$ and $D$ are the
ferromagnetic and DM exchange energies, respectively. The Zeeman
field $\v B$ is typically applied perpendicular to the planar
direction, $\v B = B\hat{z}$. The dynamics of the conduction
electrons and the coupling of their spins to the local moment, on the
other hand, are embodied in the $sd$ Hamiltonian,

\ba H_{sd} = \int d^3 \v r ~  \Psi^\dag \left( {\v p^2 \over 2m} -
J_H \bm \sigma \cdot \v n \right) \Psi, ~~ \Psi = \bpm c_{\uparrow} \\
c_{\downarrow} \epm . \label{eq:H-sd}\ea
Electron operators are given in the spinor form $\Psi$ comprising up
($c_\uparrow$) and down ($c_\downarrow$) spin components. Derivation
of the subsequent equation of motion of $\v n$ based on the total
Hamiltonian $H=H_{\v n}+H_{sd}$ is well-documented in the
literature\cite{tatara} and reproduced briefly here.

In the large Hund's coupling limit ($J_H \rightarrow \infty$) the
spin $\Psi^\dag \bm \sigma \Psi$ of the conduction electron is
forced to align with the local magnetization direction $\v n$ while
the electron band with anti-parallel spins forms a higher-energy
continuum. The idea is implemented by making the unitary
transformation, $U^\dag \v n \cdot \bm \sigma U = \sigma_z$, where
$U$ is given by

\ba U = \bpm z_1 & z_2^* \\ z_2 & -z_1^* \epm, ~ z_1 = \cos {\theta
\over 2}, ~ z_2 = e^{i\phi} \sin {\theta \over 2} . \ea
The spinor $\v z = \bpm z_1 \\ z_2 \epm$ forms the CP$^1$
representation of the classical spin $\v n = \v z^\dag \bm \sigma \v
z = (\sin \theta \cos \phi, \sin \theta \sin \phi, \cos \theta )$.
The new electron spinor $U^\dag \Psi$ has the upper (lower)
component parallel (anti-parallel) to $\v n$, out of which we choose
to keep the upper one only, denoted $\psi$. The field operator
$\psi$ gives the electron whose spin direction is projected strictly
parallel to the local moment $\v n$. The Hamiltonian for $\psi$ is

\ba H'_{sd} =\int d^3 \v r~ \psi^\dag {[ \v p + \hbar \v a ]^2 \over
2m} \psi , \ea
where the gauge field $\v a$ is derived from $\v z$ as $a_i = -i \v
z^\dag \partial_i \v z$. It is gauge-invariant under $\v z
\rightarrow e^{i\chi} \v z$, \textit{and} $\psi \rightarrow
e^{-i\chi} \psi$. Expanding the $sd$ Hamiltonian gives the coupling
of local moment to the electrons as\cite{tatara}

\ba H_{\v a - \v j} = \int d^3 \v r~  \left( \hbar \v a \cdot \v j +
{\hbar^2 \rho \over 2m} \v a^2 \right) , \label{eq:a-j-coupling}\ea
where we also introduced paramagnetic spin current $\v j$ and the
spin density $\rho$ as\cite{tatara}

\ba \v j &=& {1\over 2m} \left( \psi^\dag [ \v p \psi ] - [\v p
\psi^\dag] \psi \right) , \nn
\rho &=& \psi^\dag \psi .  \ea

We focus on the dynamics of the local moments using $H_{\v n} +
H_{\v a - \v j}$ as the total Hamiltonian and ask how an externally
imposed spin current pattern $\v j$ influences the magnetic
dynamics. Already the likelihood of Skyrmion induction by the
circulating $\v j$ can be seen in the following simple argument. In
a steady state, the divergence of the current vanishes, $\bm \nabla
\cdot \v j = 0$, making it possible to re-write $\v j$ as the curl
$\v j = \bm \nabla \times \v c$. For circulating current of
cylindrically symmetric form, the vector field $\v c$ is directed in
the $z$-direction. Integrating the energy functional by parts, one
can re-write $\v a \cdot \v j \sim - \v c \cdot \bm \nabla \times \v
a$ and find the effective magnetic field $(\bm \nabla \times \v a
)_z$ couples linearly to the current source $\v c$. This effective
magnetic field is nothing other than the Skyrmion density through
the relation

\ba (\bm \nabla \times \v a )_z = {1\over 2 } \v n \cdot (\partial_x
\v n \times \partial_y \v n). \ea
A suitable $\v c$ ($\v j$) of sufficient strength will overcome the
nucleation energy cost and induce a Skyrmion with $(\bm \nabla
\times \v a )_z \neq 0$. The sign dictates that the creation of
Skyrmion, $Q>0$, is energetically preferred for counter-clockwise
(CCW) direction of spin current, while anti-Skyrmions, $Q<0$, are
preferred for clockwise (CW) spin current direction. Here

\ba Q={ 1 \over 4\pi}\int d^2 \v r ~\v n \cdot (\partial_x \v n
\times \partial_y \v n) \ea
is the total Skyrmion charge.

The real-time action for the magnetic dynamics is

\ba {\hbar S \over a^3}  \int d^3 \v r dt ~ (1-\cos \theta )
\partial_t \phi - \int dt \bigl( H_{\v n} + H_{\v a - \v j} \bigr) \ea
with the effective moment $S$, and $( \theta, \phi )$ are the polar
and azimuthal angles of the magnetization unit vector $\v n$. The
LLG equation follows as

\ba &&  \dot{\v n}  + {1\over \hbar} \v n \times \left( -Ja^2  \bm
\nabla^2 \v n + 2  D a \bm \nabla \times \v n -  \v B \right)   \nn
&& ~~~~ + \alpha  \v n \times \dot{\v n}+ {a^3 \over S}  (\v j \cdot
\bm \nabla) \v n = 0 . \label{eq:continuum-LLG}\ea
The Gilbert damping constant $\alpha$ is introduced
phenomenologically\cite{tatara}. Assuming all the spins sharing the
same $(x,y)$-coordinate behave identically, we arrive at a
two-dimensional discretized version of the LLG equation

\ba && \dot{\v n}_i +\v n_i \times \sum_{j \in i} \Bigl( - \v n_{i+a
\hat{e}_{ji}} + \kappa \v n_{i+a \hat{e}_{ji}} \times \hat{e}_{ji}
-\kappa^2 \v b  \Bigr) \nn
&& ~~~~~~~~~~~~~~~~~~~~ + \alpha \v n_i \times \dot{\v n}_i  \nn
&& + {1\over v_0 } \Bigl(  v^x_{i}(\v n_{i+a\hat{x}}-\v n_{i} ) +
v^y_{i}(\v n_{i+a\hat{y}}-\v n_{i} ) \Bigr) = 0.
\label{eq:discrete-LLG} \ea
The drift velocity $\v v_i = (v^x_i, v^y_i)$ is related to the
current density through $a^3 \v j= p x \v v$, where $x$ is the
number of conduction electrons enclosed in a unit cell of volume
$a^3$ and $p$ is the spin polarization fraction. All physical
quantities in the equation are made dimensionless by choosing the
unit of time $t_0=\hbar/J$, and $\kappa = D/J$, $\v b=\v B
/(D^2/J)$. The quantity $v_0 = a S / t_0 px$ with the dimension of
velocity is introduced as well. The wavelength of the spiral is
given by $\lambda \sim 2\pi\sqrt{2} a/\kappa$ in two dimensions.
Square lattice of size $L\times L$ is adopted throughout the
simulation and the $\kappa$ value equal to $0.5$, which yields the
spiral wavelength $\lambda \sim 4\pi\sqrt{2}a \sim 17.8 a$. The
diameter of the Skyrmion is roughly $\lambda/2$, which makes the
Skyrmion radius $R_s \sim \lambda/4$. The sum $\sum_{j\in i}$ spans
the four nearest neighbors of the site $i$, connected by the unit
vectors $\hat{e}_{ji}$.

Time-integration of the discrete LLG equation
(\ref{eq:discrete-LLG}) is done using the fourth-order Runge-Kutta
method. Each unit of time $t_0$ is sliced into over 1000 steps for
the integration procedure, with no difference in the result found
for finer step sizes. Typical integration ran up to $t/t_0 \sim 10^3
$ without a noticeable error accumulation. The spin current density
$\v j = \v v / v_0$ is given the circulating profile with the
Lorentzian shape

\ba \v j (\v r) = \hat{\phi} j_0 { R_c^{2} \over R_c^{2} + r^{2}}
\label{eq:j-profile}\ea
at the center of the simulation lattice, characterized by its
magnitude $j_0$ and the extent $R_c$. Both positive and negative
$j_0$ corresponding to the CCW (CW) circulation of the spin current
were considered. In our convention, a uniform spin current $\v j =
\v j_0$ leads to the Skyrmion flow velocity $\v v_s$ in the same
direction, $\v v_s \parallel \v j_0$, opposite to what was used in
Ref. \onlinecite{zang}. Previous study\cite{bogdanov,han,tokura}
found the magnetic phase with helical spin configuration at zero
magnetic field and Skyrmion crystal phase over the intermediate
field values $b_{c1}\lesssim b \lesssim b_{c2}$. Two magnetic fields
are considered in our simulation here, one for $b$ slightly less
than the upper critical field strength $b_{c2}$ where a typical
ground state consists of ferromagnetic spins dotted with a few
isolated anti-Skyrmions (Fig.
\ref{fig:FM-spin-creation}b)\cite{tokura,FeGe,han}, and the other at
$b=0$ and having the helical spins as the ground state (Fig.
\ref{fig:helical-spin-creation}b). We refer to these two ground
states as FM$'$ (FM+anti-Skyrmions) and H, respectively.

\begin{widetext}
\begin{center}
\begin{figure}[ht]
\includegraphics[width=150mm]{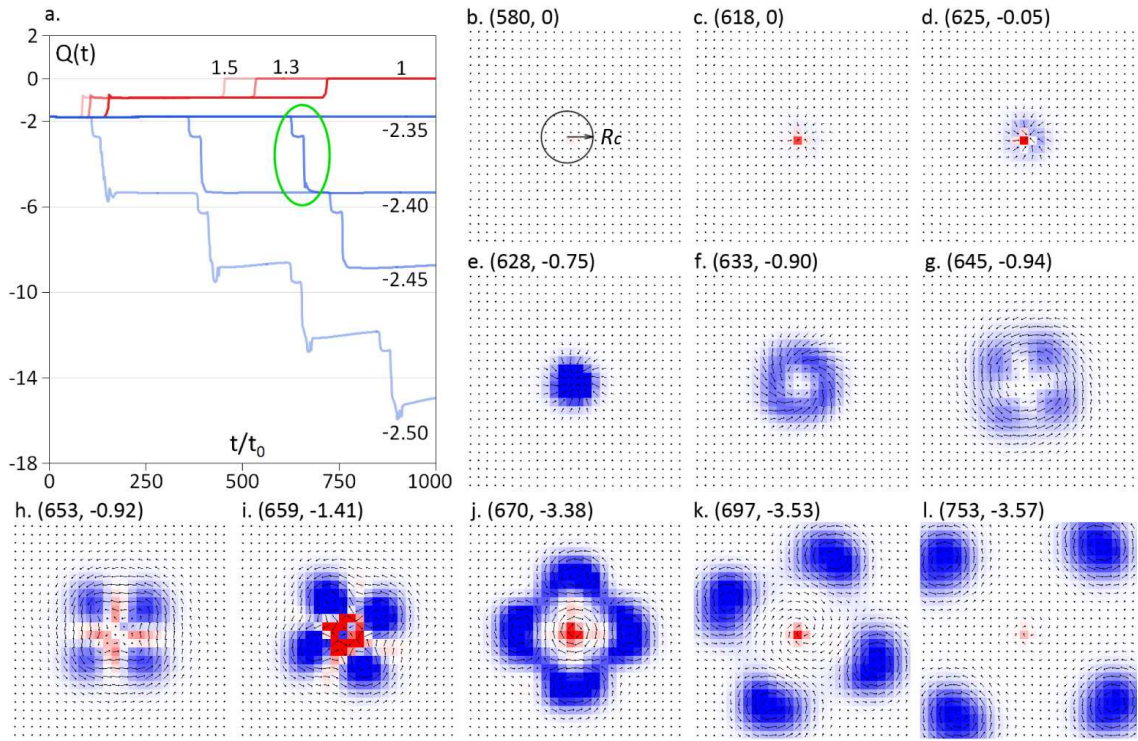}
\caption{(color online) Skyrmion generation by circulating spin
current source in the FM$'$ background. (a) Time dependence of the
total Skyrmion number $Q(t)$. $Q(0)\approx -2$ is nonzero from the
residual anti-Skyrmions in the ferromagnetic background. Blue
(red)-colored curves correspond to CW (CCW) circulating current with
the $j_0$ value indicated for each curve. (b)-(l) Time-dependent
snapshots of the spin configuration over the green-circled time
interval in (a). Adjacent spins are separated by the distance $a$.
The circle at the center of (b) indicates $R_c$, the extent of the
current source. Residual anti-Skyrmions lie outside the field of
view. Red and blue refer to positive and negative Skyrmion
densities, respectively. Time at which the snapshot is taken and the
change in the Skyrmion number are given as $(t/t_0, Q(t)-Q(0))$
above each figure. Skyrmion number is seen to decrease by one, and
later by three. Threshold value $j_{0c}$ lies between 2.35 and 2.4
for CW spin current.}\label{fig:FM-spin-creation}
\end{figure}
\end{center}
\end{widetext}

\begin{widetext}
\begin{center}
\begin{figure}[ht]
\includegraphics[width=150mm]{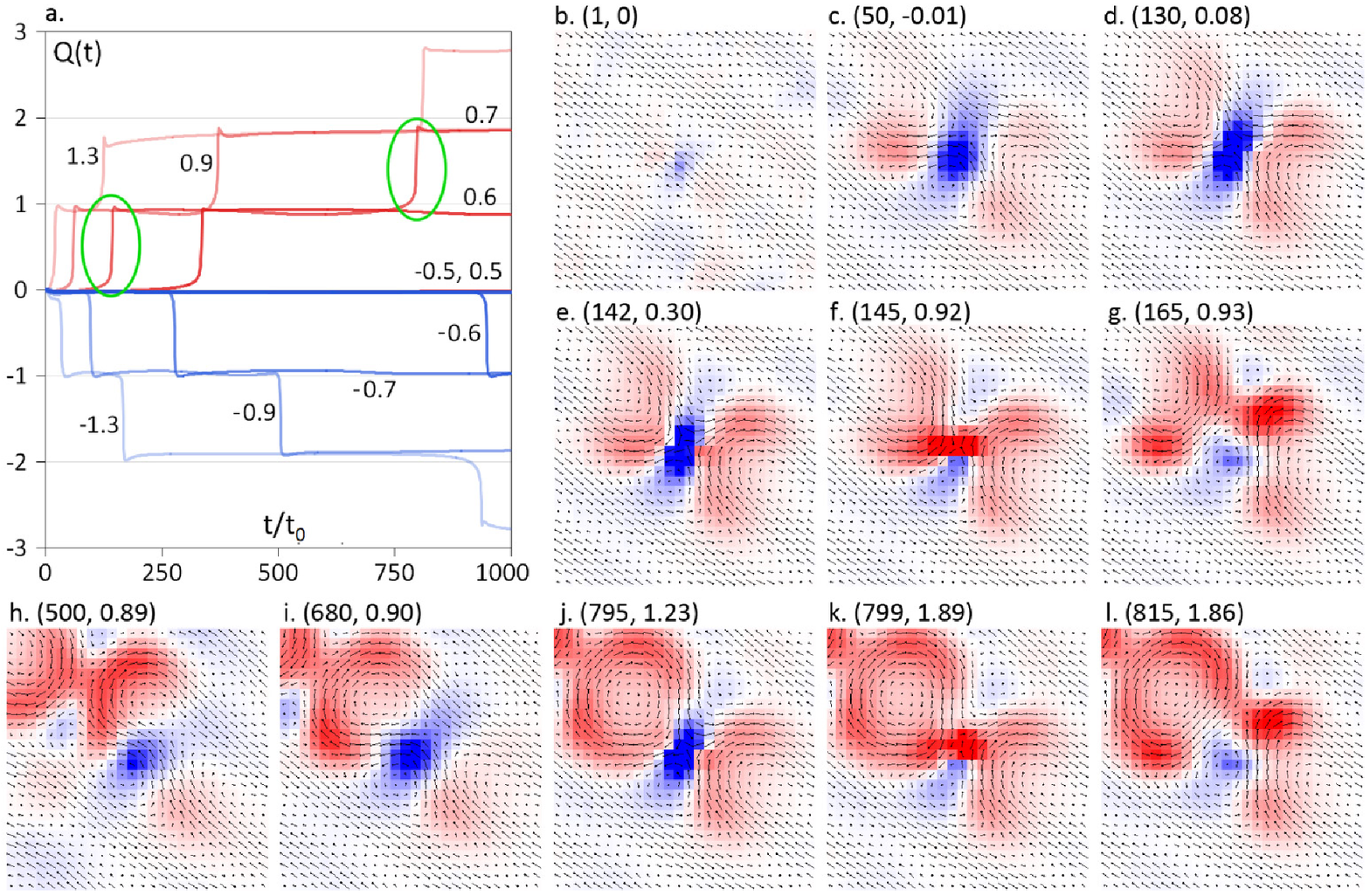}
\caption{(color online) Skyrmion generation by circulating spin
current source in the H background. (a) Time dependence of the total
Skyrmion number $Q(t)$ for various $j_0$ values. Initial Skyrmion
number is zero for the helical spin. Same color convention as in
Fig. 1 applies throughout this figure. (b)-(l) Two sets of snapshots
corresponding to $Q=0\rightarrow 1$ and $Q=1\rightarrow 2$
corresponding to two green circles in Fig. 2a.}
\label{fig:helical-spin-creation}
\end{figure}
\end{center}
\end{widetext}

\section{Processes of current-induced Skyrmion formation}
\label{sec:process}

\subsection{Skyrmion production}
Due to the simpleness of the ferromagnetic spin background, it is
easier to understand the formation process of Skyrmions for the
FM$'$ case, $b\lesssim b_{c2}$. For optimal Skyrmion generation
condition we choose $R_c$ comparable to the typical radius of a
Skyrmion, $R_c \approx R_s$. As the system evolves over time, the
time-dependent Skyrmion number $Q(t)= (1/4\pi) \int d^2 \v r ~ \v n
(t) \cdot (\partial_x \v n (t) \times\partial_y \v n (t) )$ is
evaluated to keep track of the Skyrmion creation process.

As shown in Fig. \ref{fig:FM-spin-creation}, CW spin current
produces, for a brief moment, a spin texture that corresponds
roughly to a Skyrmion-anti-Skyrmion pair within the radius
$r\lesssim R_s$ (red core region in Fig.
\ref{fig:FM-spin-creation}c-d). Out of the pair, the Skyrmion part
soon vanishes, leaving behind an anti-Skyrmion (Fig.
\ref{fig:FM-spin-creation}e-g). The anti-Skyrmion then gradually
drifts out, due to the finite Gilbert damping which gives rise to a
velocity orthogonal to the local current flow. The Gilbert
damping-induced Skyrmion Hall motion was extensively discussed in
Ref.~\onlinecite{zang} to which we refer the interested readers. The
Skyrmion meanwhile experiences an inward flow and vanishes. Once the
anti-Skyrmion has drifted a sufficient distance out, another burst
of Skyrmion-anti-Skyrmion pair takes place (Fig. 1h), with the
Skyrmion portion vanishing through the core again decreasing the
total Skyrmion number by an integer amount (Fig. 1i-l). Each burst
of the Skyrmion-anti-Skyrmion pair and the subsequent decay of the
Skyrmion is responsible for the total Skyrmion number $Q(t)$ jumping
by an integer amount as shown in Fig. \ref{fig:FM-spin-creation}a.

Evidently creation of Skyrmion-anti-Skyrmion pair (Skyrmion dipole)
is energetically cheaper  than creating an isolated Skyrmion
(Skyrmion monopole). Once a pair is created, the Gilbert
damping-induced Hall motion naturally moves the two objects in the
opposite directions, so that only one species of Skyrmions can leave
the current core. When CCW current $(j_0 > 0)$ is applied, the same
Gilbert drifting mechanism that pushed the anti-Skyrmions out now
tends to absorb the nearby anti-Skyrmions toward the core region
because the direction of the Hall motion is also reversed. When an
anti-Skyrmion is pulled sufficiently close to the center, a brief
burst of Skyrmion appears at the core and ``eats up" the attracted
anti-Skyrmion. A good mental picture of the process is to take Fig.
1b-1g and reverse them in time. When all the existing anti-Skyrmions
are eaten up in this way, the total Skyrmion number $Q(t)$ remains
close to zero (Fig. \ref{fig:FM-spin-creation}a) and does not rise
above it as a consequence of the fact that positive charge Skyrmions
are energetically forbidden for the magnetic field $b>0$.

There is a critical value of the spin current density $j_{0c}$
required to produce anti-Skyrmions out of the ferromagnetic spin
background. Obviously the rate of energy input from the circulating
current needs to outpace that of the energy drain due to the Gilbert
damping in order to supply sufficient energy to create an
anti-Skyrmion. Above the threshold, $| j_0 |
> j_{0c}$, the total Skyrmion number begins to jump in integer steps with
time intervals that decrease as $| j_0 | - j_{0c}$ increases
(Fig.~\ref{fig:FM-spin-creation}a).

Figure \ref{fig:helical-spin-creation} depicts the Skyrmion creation
process for H background $(b=0)$. As with the FM$'$ background,
$Q(t)$ can be seen to increase in integer quantized steps. In
contrast to $b>0$ case which prefers an anti-Skyrmion charge,
opposite signs of the circular current result in more or less
symmetrical profiles of $Q(t)$ as shown in Fig.
\ref{fig:helical-spin-creation}a. Snapshots of spin configurations
for $Q=0 \rightarrow Q=1$ are in Fig. 2b-g, and those of $Q=1
\rightarrow Q=2$ in Fig.
\ref{fig:helical-spin-creation}h-\ref{fig:helical-spin-creation}l.
At first a small patch of helical spin is torn into two disjoint
segments as in Fig. \ref{fig:helical-spin-creation}c-d near the
current core, giving way to the nucleation of some defects in the
intervening region. The defect profile is best described as the
meron-anti-Skyrmion-meron composite (merons are in blue,
ant-Skyrmion in red in Fig.~\ref{fig:helical-spin-creation}c), out
of which the central anti-Skyrmion shrinks in size while the two
merons drift out, again due to the Gilbert damping-induced Hall
motion. The two merons are responsible for the total charge $Q(t)
\approx 1$. Each time the process repeats itself, the Skyrmion
number increases by one.

In both H and FM$'$ spin backgrounds, the resulting charge of the
Skyrmions produced by the current is correlated with the helicity of
the circulation as predicted in an earlier argument; see paragraph
following Eq. (6). In detail, however, the production process proves
to be far more intricate involving an Skyrmion-anti-Skyrmion pair at
the initial stage of metamorphosis rather than the simple creation
of an isolated Skyrmion out of the vacuum. A different mechanism
then intervenes, namely Gilbert damping-assisted Hall motion, which
separates the pair according to their respective charges, leaving
only one species of Skyrmions to survive at the end.
\\

\begin{figure}[ht]
\includegraphics[width=85mm]{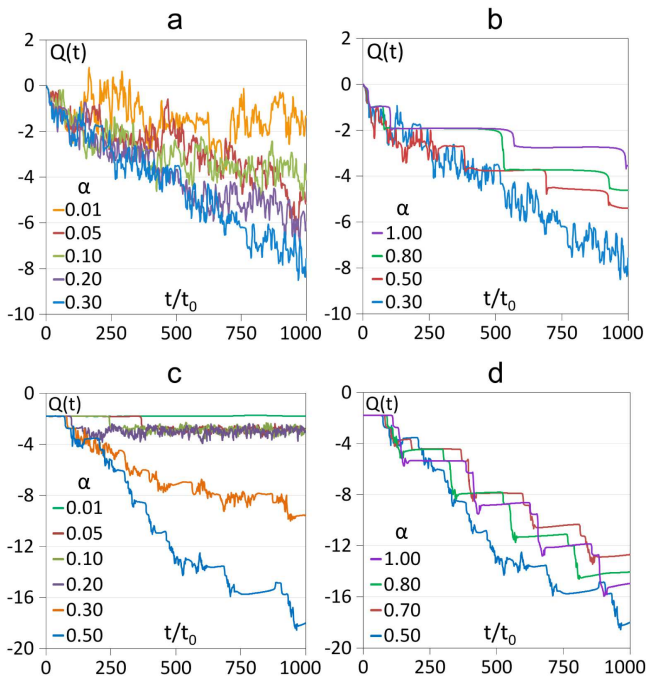}
\caption{(color online) Influence of Gilbert damping on Skyrmion
generation. (a)-(b) Plot of Skyrmion number $Q(t)$ with $b=0$ (H
background) for different Gilbert damping constants $\alpha$.
$|Q(t)|$ increases monotonically with $\alpha$ without showing signs
of quantization when $\alpha$ is small. At a larger damping, $0.3 <
\alpha < 0.5$, quantization of Skyrmion number begins to occur. The
time span of each integer plateau becomes longer, and the production
rate of Skyrmions becomes less, with increasing $\alpha$. (c)-(d)
Plot of the Skyrmion number $Q(t)$ with FM$'$ background. Numbers in
the inset are the Gilbert damping constants. $(j_0, R_c) = (-1.5,
3.0)$ was used for (a)-(b) and $(j_0, R_c ) = (-2.5, 3.0)$  for
(c)-(d). }\label{fig:Gilbert}
\end{figure}

\subsection{Influence of Gilbert damping}

Gilbert damping plays a major role in the generation of Skyrmions
and anti-Skyrmions by facilitating their radially outward motion
from the circulating spin current region. All results shown in
Figures \ref{fig:FM-spin-creation} and
\ref{fig:helical-spin-creation} are obtained for $\alpha = 1$, which
is a fairly large value for Gilbert damping parameter. As the
damping is reduced, the sharp, integer quantization of Skyrmion
number $Q(t)$ gets less conspicuous as shown in Fig.
\ref{fig:Gilbert}. According to our numerical observation, clear
integer quantization of $Q(t)$ is more or less synonymous with the
production of well-isolated Skyrmions in real space, which only
becomes possible for $\alpha$ in excess of certain minimum value as
shown in Fig. \ref{fig:Gilbert}b and \ref{fig:Gilbert}d. This is
related to the quite complex manner in which the Skyrmion creation
process takes place. The initial creation process, as shown in Figs.
\ref{fig:FM-spin-creation} and \ref{fig:helical-spin-creation},
always involves a pair of Skyrmions of opposite charges, out of
which only one is pushed outwards by virtue of Gilbert
damping-assisted Hall motion. A large Hall motion requires a
correspondingly large Gilbert damping constant. Otherwise, when
$\alpha$ is small, those Skyrmions after generation continue to move
in circular path (because current is circular) rather than drifting
out, and collide/merge with other Skyrmions. This leads to a large
overlap between spatially adjacent Skyrmions, as well as a
non-integral value of $Q(t)$\cite{comment}. Only with sufficiently
large $\alpha$ do we find that the Skyrmions are being pushed out
fast enough after their generation to minimize the overlap and give
rise to integer $Q(t)$. Mochizuki's recent LLG
simulation\cite{mochizuki} uses the Gilbert constant $\alpha \leq
0.04$, while extra damping effects due to spin-motive force are
expected to increase this value to around 0.1\cite{zang}. Further
enhancing the Gilbert damping might require intentional disordering
of the film by irradiation.

\subsection{Skyrmion lifetime}

\begin{figure}[ht]
\includegraphics[width=85mm]{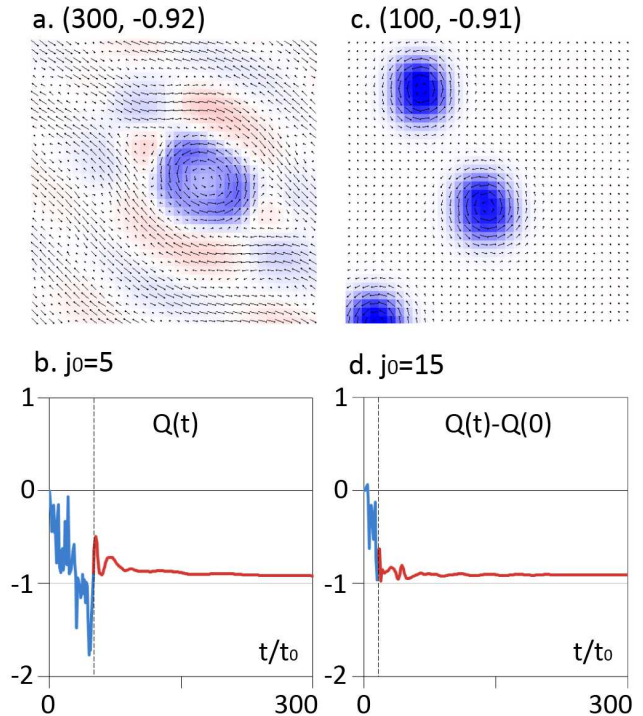}
\caption{(color online) Skyrmion generation by pulse current.
(a)-(b) Pulse current of size $j_0 = 5$ is applied for $ 0 < t_1/t_0
< 50$ over the helical spin state and then turned off. (c)-(d) $j_0
= 15$ for $0 < t_1 / t_0 < 15$ over the FM$'$ state. Dotted vertical
line in (b) and (d) indicates the termination of the applied
current. Top row: the spin configuration and Skyrmion number
$Q(t)-Q(0)$ when a sufficiently long time has elapsed after the
current is turned off. Bottom row: time dependence of the Skyrmion
number $Q(t)-Q(0)$. All current directions are CCW with $R_c = 3$
and $\alpha = 0.1$.}\label{fig:pulse}
\end{figure}

Once the Skyrmions have been generated by means of current, it is
important that they remain there after the current is turned off if
they are to serve as memory bits. After the current was applied long
enough to nucleate a Skyrmion, we turned it off in the simulation to
see if it would decay. In fact, the decay of Skyrmion is expected
because without the current, the Skyrmion would be in a metastable
state. As the $Q(t)$ plots in Fig. \ref{fig:pulse}b and
\ref{fig:pulse}d show, however, the anti-Skyrmions once created
remain extremely stable although energetically it cannot exist in
the ground state for $b=0$. As far as our numerical simulation
lasts, the spin patterns shown in Fig. \ref{fig:pulse}a and
\ref{fig:pulse}c remained persistently the same. Such a long
lifetime of an isolated Skyrmion is a positive feature in utilizing
the chiral ferromagnet as a platform for encoding Skyrmion-based
information.

\subsection{Effects of radial current pulse}

Earlier we used the circulating current as a platform to generate
Skyrmions electrically. Clear, integer quantization of Skyrmion
charge was possible only for very large Gilbert damping parameters.
Here we show that Skyrmions can also be created by radial current
pulses as given by

\ba \v j (\v r) = \hat{ r } { j_0  \over \sqrt 2 r } .
\label{eq:j-profile}\ea

\begin{figure}[ht]
\includegraphics[width=85mm]{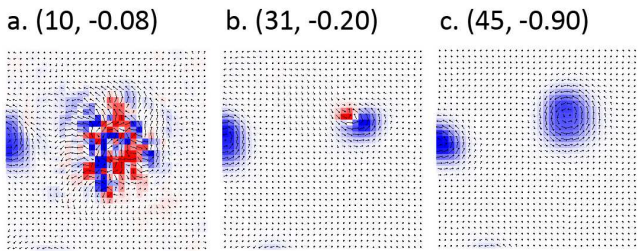}
\caption{(color online) Skyrmion creation by radial current source.
Current pulse with $j_0=10$ lasting for $t/t_0=10$ was given. The
numbers above each figure indicate $(t/t_0, Q(t)-Q(0))$. (a) Right
at the turn-off, one finds a swarm of positive and negative Skyrmion
fragments. (b) After some time has passed, most of the Skyrmions and
anti-Skyrmions have annihilated each other, leaving behind a
predominant anti-Skyrmion density due to the external magnetic
field. (c) Final state shows one anti-Skyrmion.}\label{fig:radial}
\end{figure}

While the radial current source remains on, a bunch of plus and
minus Skyrmion fragments appear from the center and drift out as
they collide and annihilate with oppositely charged fragments as
shown in Fig. \ref{fig:radial}a. When the radial current is turned
off, most of the plus and minus Skyrmion fragments annihilate,
leaving behind a portion that evolves into an integer-valued
Skyrmion, with the sign as favored by the direction of the applied
magnetic field, as shown in Fig. \ref{fig:radial}b and
\ref{fig:radial}c. This way of creating Skyrmion is only possible
when external magnetic field is acting on the helical magnet. For
helical spin configuration, the plus and minus Skyrmion densities
created by the radial current annihilate and disappear completely,
returning back to the initial helical spin configuration over time.

We emphasize that this method of generating Skyrmions in the FM$'$
background works very well even for small values of the Gilbert
damping constant $\alpha$. This is because the selection process to
isolate one particular sign of Skyrmion is done by external magnetic
field, not by Gilbert damping as in the case of circulating current.

\section{Experimental realization and discussion}
\label{sec:exp}

The typical exchange energy scale of a chiral ferromagnet is
$J_0\sim 3$ meV\cite{grigoriev}, with the dispersion $\hbar\omega_k
\sim J_0 (k a_0)^2$ involving $a_0$, the linear dimension of the
physical unit cell. On the other hand, from our continuum LLG
equation in (\ref{eq:continuum-LLG}) follows the dispersion relation
$\hbar \omega_k \sim J (k a)^2$. Hence, $J$ used in our model is
related to the microscopic parameter $J_0$ by $J \sim J_0 (a_0
/a)^2$, which also sets the time and the velocity scales as $t_0 =
\hbar/J \approx 220(a / a_0)^2$ fs and $v_0 = a S/t_0 px \sim (a_0
J_0 / \hbar) ( S /p) (a_0 / a)^4\sim 1.8 \times 10^3 (S/p) (a_0
/a)^4$ m/s.

In order to fix the ratio $a/a_0$, we note that with our choice
$\kappa = 0.5$ the typical wavelength of the spiral is $\lambda =
4\pi \sqrt{2} a \approx 17.8 a$. Values of $\lambda$ vary quite
markedly among the chiral ferromagnets, ranging from $\sim$3 nm for
FeGe\cite{MnGe}, $\sim$20 nm for MnSi\cite{pfleiderer1}, $\sim$70 nm
for FeGe\cite{FeGe}, and $\sim$90 nm for
Fe$_{1-x}$Co$_x$Si\cite{pfleiderer2,tokura}. Taking the
characteristic spiral wavelength $\lambda \sim 60$ nm yields $17.8 a
\sim 60$ nm, or $a\sim 4$ nm. Choosing the microscopic lattice
constant $a_0 = 4$\AA~ then gives the ratio $a/a_0 \sim 10$.
Finally, with the spin polarization fraction $p\sim 0.1$ and $S=1$
we obtain $ v_0 \sim 1$ m/s. The critical current density $j_{0c}$
of order unity we saw in the simulation thus corresponds to the
drift velocity of a few m/s to be imparted to the conduction
electrons. Estimate of the carrier concentration in MnSi\cite{ong}
$\rho_e \sim 10^{28}-10^{29}$/m$^3$ yields the critical current
density $j_c \sim 10^{10}-10^{11}$A/m$^2$, which is close to, or
below the critical current required for domain wall switching in
typical magnetic devices\cite{tatara}. The time interval between
successive Skyrmion creation events as shown in Figs.
\ref{fig:FM-spin-creation} and \ref{fig:helical-spin-creation} is
$\sim 100t_0$. With the estimate $t_0 \sim 20$ ps, the time interval
is $\sim 2$ ns. The requirement for Skyrmion nucleation is reduced
to having a strong current pulse of a few nanosecond duration with
the density $j_c \sim 10^{10}-10^{11}$A/m$^2$ - a practice readily
available in modern-day spintronics laboratories.

Although a detailed practical scheme to implement the circuitry
needed for the Skyrmion generation is yet to be worked out, our
simulation provides a proof-of-concept demonstration for Skyrmion
creation by electrical current. It is conceivable, in principle,
that an artificial bend in the current pathway (like a curved
freeway) is fabricated to mimic the circular pattern we assumed in
the simulation. As the current of enough intensity passes through
such a bend, Skyrmions will ``pop out" from the point of greatest
curvature. The Skyrmion production by means of radial current
discussed in Sec. III D can be mimicked by designing a narrow
channel of metallic magnet that opens up into a much wider region. A
current coming through such constriction will be a source of radial
motion, thus of Skyrmions. Although much engineering ideas need to
be carved out yet, the fundamental processes revealed by the present
numerical simulation will be exciting to realize in laboratories in
the near future.

\acknowledgments This work is supported by Mid-career Researcher
Program (No. 2010-0008529). Valuable comments on the manuscript from
Suk-Bong Choe and discussions with Hyun-Woo Lee are acknowledged. We
also thank professor Naoto Nagaosa for careful reading of the
initial manuscript and suggestions.

%\bibitem{rasing-nature}  Kimel, A. V., Kirilyuk, A.
%Usachev, P. A., Pisarev, R. V., Balbashov, A. M. \& Rasing, Th.
%Ultrafast non-thermal control of magnetization by instantaneous
%photomagnetic pulses. \textit{Nature} \textbf{435}, 655-657 (2005).
%
%\bibitem{rasing-PRL} Stanciu, C. D., Hansteen F., Kimel A. V., Kirilyuk A.,
%Tsukamoto, A.,  Itoh, A. \&  Rasing Th. All-Optical magnetic
%recording with circularly polarized light. \textit{Phys. Rev. Lett.}
%\textbf{99}, 047601 (2007).
%
%\bibitem{tokura-APL} Ogasawara, T., Iwata, N., Murakami, Y.,
%Okamoto, H. \& Tokura, Y. Submicron-scale spatial feature of
%ultrafast photoinduced magnetization reversal in TbFeCo thin film.
%\textit{App. Phys. Lett.} \textbf{94}, 162507 (2009).


\begin{thebibliography}{99}
\bibitem{comment} In a continuum theory one should obtain integer
$Q(t)$ even for overlapping Skyrmions. The non-integral values are a
result of the calculation being carried out on a discrete lattice.
\end{thebibliography}
\end{document}